\pdfoutput=1
\documentclass[aps,prl,twocolumn,superscriptaddress,preprintnumbers,floatfix]{revtex4-1}

\usepackage{graphicx}
\usepackage{amsmath}
\usepackage[caption=false]{subfig}
\usepackage{siunitx}
\usepackage{placeins}
\usepackage{color}
\usepackage{standalone}
\usepackage{dcolumn}
\usepackage{tensor}
\usepackage{bm}
\usepackage{microtype}
\usepackage{etoolbox}
\usepackage{amssymb}
\usepackage{mathrsfs}
\usepackage{accents}
\usepackage[normalem]{ulem}
\usepackage[dvipsnames]{xcolor}
\usepackage[colorlinks,urlcolor=NavyBlue,citecolor=NavyBlue,linkcolor=NavyBlue,pdfusetitle]{hyperref}
\usepackage[all]{hypcap}
\usepackage[inline]{enumitem}
\usepackage[utf8]{inputenc}
\usepackage{lipsum}
\usepackage{booktabs}
\usepackage{wasysym}

\usepackage{array}
\newcolumntype{L}[1]{>{\raggedright\let\newline\\\arraybackslash\hspace{0pt}}m{#1}}
\newcolumntype{C}[1]{>{\centering\let\newline\\\arraybackslash\hspace{0pt}}m{#1}}
\newcolumntype{R}[1]{>{\raggedleft\let\newline\\\arraybackslash\hspace{0pt}}m{#1}}

\newcommand{\MKI}{\affiliation{Department of Physics and Kavli Institute for Astrophysics and Space Research, Massachusetts Institute of Technology, 77 Massachusetts Ave, Cambridge, MA 02139, USA}}

\newcommand{\heading}[1]{\textbf{\textit{#1.}}}

\newcommand{\beq}{\begin{equation}}
\newcommand{\eeq}{\end{equation}}

\newtoggle{commentsoff}
\togglefalse{commentsoff}

\ifdefined\nocomments
    \toggletrue{commentsoff}
\fi

\iftoggle{commentsoff}{
  \newcommand*{\SB}[1]{}
  \newcommand*{\MI}[1]{}
  \newcommand*{\sv}[1]{}
  \newcommand*{\VV}[1]{}
  \newcommand*{\comment}[1]{}
  
  \newcommand*{\todo}[1]{}
  \newcommand*{\warn}[1]{}

}{
  \newcommand*{\MI}[1]{{\color{magenta} [{\bf MAX}: #1]}}
  \newcommand*{\SB}[1]{{\color{RedOrange} [{\bf SYLVIA}: #1]}}
  \newcommand*{\sv}[1]{\textcolor{ForestGreen}{[\textbf{SALVO}: #1]}}
  \newcommand*{\VV}[1]{\textcolor{Purple}{\textbf{VIJAY}: #1}}
  
  \newcommand*{\comment}[1]{{\color{blue} [{\bf NOTE}: #1]}}
  \newcommand*{\warn}[1]{{\color{red} [{\bf WARNING}: #1]}}
  \newcommand*{\todo}[1]{{\color{red} [{\bf TODO}: #1]}}

}

\graphicspath{{./fig/}}

\begin{document}


\title{A new spin on LIGO-Virgo binary black holes}

\author{Sylvia Biscoveanu}
\email[]{sbisco@mit.edu}
\affiliation{
LIGO Laboratory, Massachusetts Institute of Technology, Cambridge, Massachusetts 02139, USA
} \MKI

\author{Maximiliano Isi}
\thanks{NHFP Einstein fellow}
\affiliation{
LIGO Laboratory, Massachusetts Institute of Technology, Cambridge, Massachusetts 02139, USA
} \MKI

\author{Salvatore Vitale}
\affiliation{
LIGO Laboratory, Massachusetts Institute of Technology, Cambridge, Massachusetts 02139, USA
} \MKI

\author{Vijay Varma}
\thanks{Klarman fellow}
\affiliation{TAPIR, California Institute of Technology, Pasadena, CA 91125,
USA}
\affiliation{Department of Physics, and Cornell Center for Astrophysics and
Planetary Science, Cornell University, Ithaca, New York 14853, USA}

\hypersetup{pdfauthor={Biscoveanu, Isi, Vitale, Varma}}

\date{\today}

\begin{abstract}
Gravitational waves from binary black holes have the potential to yield information on both of the intrinsic parameters that characterize the compact objects: their masses and spins. While the component masses are usually resolvable, the component spins have proven difficult to measure.
This limitation stems in great part from our choice to inquire about the spins of the most and least massive objects in each binary, a question that becomes ill-defined when the masses are equal.
In this paper we show that one can ask a different question of the data: what are the spins of the objects with the highest and lowest dimensionless spins in the binary? We show that this can significantly improve estimates of the individual spins, especially for binary systems with comparable masses.
When applying this parameterization to the first 13 gravitational-wave events detected by the LIGO-Virgo collaboration (LVC), we find that the highest-spinning object is constrained to have nonzero spin for most sources and to have significant support at the Kerr limit for GW151226 and GW170729. 
A joint analysis of all the confident binary black hole detections by the LVC finds that, unlike with the traditional parametrization, the distribution of spin magnitude for the highest-spinning object has negligible support at zero spin.
Regardless of the parameterization used, the configuration where all of the spins in the population are aligned with the orbital angular momentum is excluded from the 90\% credible interval for the first ten events and from the 99\% credible interval for all current confident detections.

\end{abstract}


\maketitle



\heading{Introduction}
Gravitational waves from compact binary coalescences (CBCs) carry imprints of the spin angular momenta $\vec{S}$ of the black holes (BHs) or neutron stars (NSs) that originated them.
The Advanced LIGO \cite{TheLIGOScientific:2014jea} and Virgo \cite{TheVirgo:2014hva} detectors can extract this information to obtain key insights about the astrophysics of compact binaries;
because the magnitude and orientation of the spins reflect the system's history, such a measurement could reveal the binary's formation mechanism~\cite{Gerosa:2013laa, Gerosa:2018wbw}.
For instance, we expect the spins of compact binaries formed in isolation to be preferentially aligned with the orbital angular momentum $\vec{L}$ \cite{Tutukov:1993, Kalogera:1999tq, Grandclement:2003ck, Postnov:2014tza, Belczynski:2016obo, Mandel:2015qlu, Marchant:2016, Rodriguez:2016vmx, OShaughnessy:2017eks, Stevenson:2017tfq}, while the same is not true of binaries formed dynamically~\cite{Sigurdsson:1993zrm, Miller:2008yw, Mandel:2009nx, Zwart:2010kx, Benacquista:2011kv, Rodriguez:2016vmx}.
Identifying the formation channel of compact binaries is one of the most pressing open problems in astrophysics, making the measurement of component spins a high-value target.

Unfortunately, the ability to measure individual spins with the LIGO and Virgo detectors has been limited, since little information about these quantities is imprinted in the inspiral waveform at leading order~\cite{Damour:2001, Racine:2008qv, Ajith:2011, Ng:2018neg}. At the population level, current inferences on the black hole spin distribution indicate that most sources have low spin magnitudes when considering the distributions of both the individual component spins~\cite{LIGOScientific:2018jsj, Kimball:2020opk} and of the spin components aligned with~\cite{Farr:2017uvj, Tiwari:2018qch, Roulet:2018jbe, Miller:2020zox} and perpendicular to~\cite{Fairhurst:2019srr} the orbital angular momentum.
%
In this paper, we show that we can draw clearer conclusions about the spins of individual objects by using a more suitable basis.
Rather than attempting to identify the spin of the heaviest and lightest of the two objects, as is usually done, we infer the properties of the objects with the highest and lowest dimensionless spin.
This straightforward reparametrization of the problem can cast a new light on the component spin measurements for near-equal-mass binaries, which appear to be the majority~\cite{Roulet:2018jbe, LIGOScientific:2018jsj, Fishbach:2019bbm}. In the following, we present our proposed reparametrization and demonstrate its impact both on simulated signals and on actual LIGO-Virgo detections.

\heading{Approach}
Within general relativity, a CBC signal is fully determined by a set of parameters encoding the intrinsic properties of the binary as well as extrinsic parameters specifying its distance and orientation.
The intrinsic parameters correspond to the mass $m_i$ and dimensionless spin $\vec{\chi}_i = \vec{S}_i c/(Gm^2)$ of each component object $i \in \{1,2\}$, plus additional quantities incorporating matter effects and eccentricity.
Virtually all of the literature, including LIGO-Virgo collaboration papers~\cite{Abbott:2019ebz, LIGOScientific:2018mvr, Abbott:2020uma, LIGOScientific:2020stg}, labels the compact objects with respect to their mass, with the index 1 corresponding to the heaviest of the two objects and 2 to the lightest, $m_1 \geq m_2$.
However, this choice is suboptimal for systems with similar masses, as it becomes degenerate for $m_1 = m_2$.
In that limit, the standard mass-based sorting induces undesired structure in the posteriors for the spin parameters.\footnote{Assuming a universal equation of state, tidal parameters for binary NSs should be unaffected, since the least massive object should be the most deformable.}

To avoid these degeneracies, we instead propose to identify objects by their dimensionless spin magnitude $\chi = |\vec{\chi}|$, and define an equivalent set of quantities $m_j$ and $\vec{\chi}_j$ for $j \in \{A,\, B\}$, with $A$ referring to the object with the highest spin and $B$ to the lowest, $\chi_A \geq \chi_{B}$.
(In the equal-mass limit, sorting by dimensionless spin is equivalent to sorting by the component angular momenta, $\vec{S}_i$.)
This amounts to a coordinate transformation effecting $\chi_{A} = \max (\chi_1, \chi_2)$ and $\chi_{B} = \min (\chi_1, \chi_2)$.
The mass of the highest-$\chi$ component is $m_A$, just as $\chi_1$ is the spin magnitude of the highest-$m$ component.
In the following, we will refer to the usual $\{1,\,2\}$ parametrization as \emph{mass sorting}, and to the new $\{A,\, B\}$ parameterization as \emph{spin sorting}.

\heading{Simulated signal}
Before we apply the spin sorting to real detections, we perform Bayesian parameter estimation on a simulated equal-mass binary BH (BBH) system with $\chi_{A}=0.8$ and $\chi_{B}=0$ to demonstrate the resolving power of the new parameterization.
The system has a redshifted total mass of $80~M_{\odot}$, and is oriented nearly edge-on with an inclination angle $\theta_{JN} = 80.21^{\circ}$.
The luminosity distance, $d_{L} = 831.47~\mathrm{Mpc}$, is chosen so that the signal is recovered with a network signal-to-noise ratio (SNR) of 30 by the two advanced LIGO instruments plus advanced Virgo, all operating at design sensitivity~\cite{TheLIGOScientific:2014jea, TheVirgo:2014hva}.
The tilt of the spinning object is $\theta_{A} = 90^{\circ}$ with respect to the orbital angular momentum, meaning that the spin vector lies entirely in the orbital plane.

\begin{figure}
	\centering
	\includegraphics[width=\linewidth]{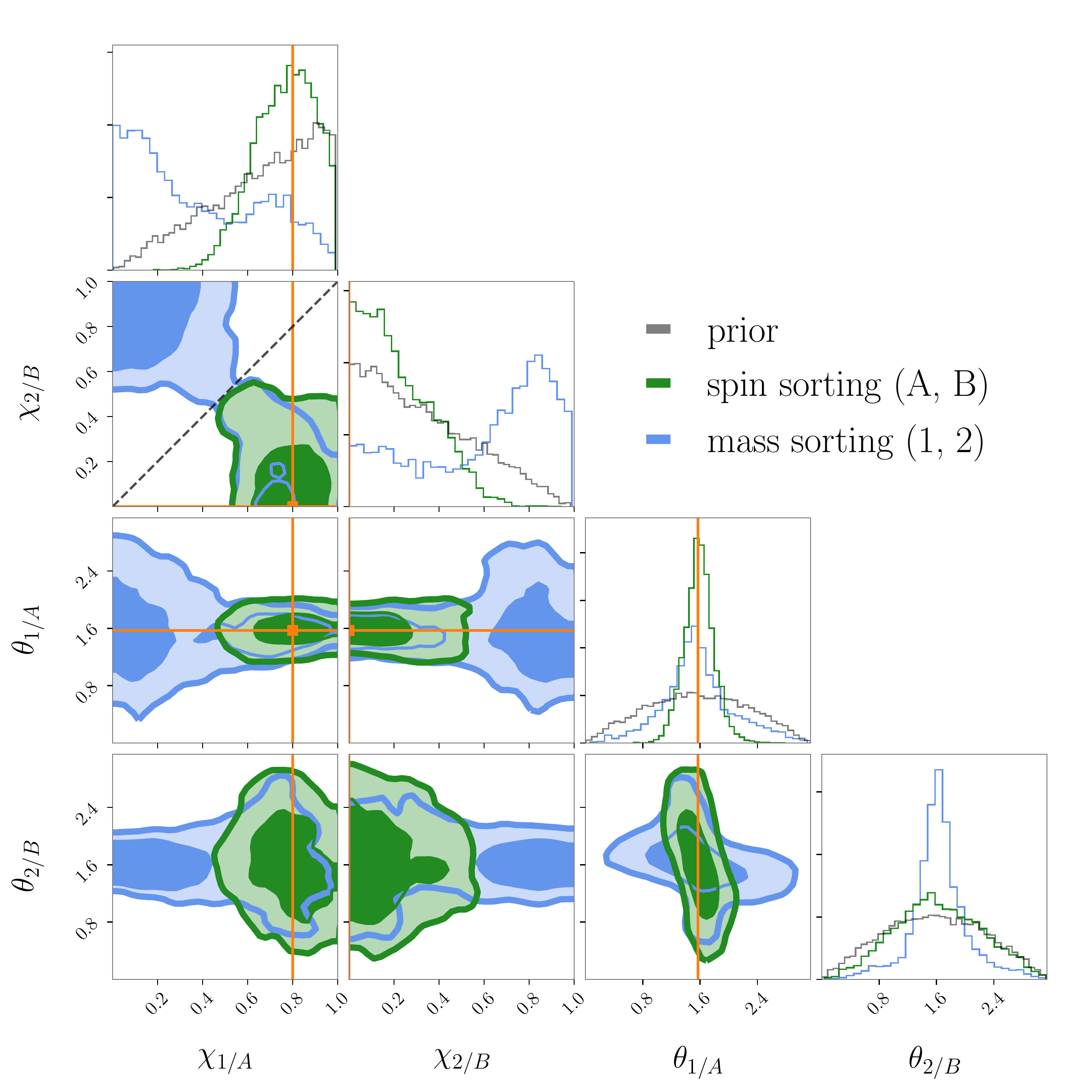}
	\caption{Comparison corner plot showing the spin magnitudes and tilts recovered for our simulated equal-mass signal using both the mass sorting in green and the spin sorting in blue. The marginalized one-dimensional priors for the spin sorting are shown in grey. Orange lines mark the true value, and the equal-spin diagonal is shown as a dashed line for reference.}
	\label{fig:injection}
\end{figure}

%
We assume standard priors for LIGO-Virgo analyses~\cite{Veitch:2014wba, LIGOScientific:2018mvr};
these imply a disjoint uniform prior on $m_A$ and $m_{B}$, and a uniform two-dimensional prior on $\chi_{A},\, \chi_{B}$.
As is the case for $m_1$  and $m_2$, the definition $\chi_{A} > \chi_{B}$ results in a ``triangular'' marginal prior for $\chi_{A}$ and $\chi_{B}$, i.e.~a probability density linearly increasing and decreasing, respectively, with the quantity (black histograms in Fig.~\ref{fig:injection}).
In order to isolate the effect of the chosen parameterization on the recovered posteriors, we do not add noise to the simulated data~\cite{Vallisneri:2007ev}. 

\begin{table}
    \centering
\caption{Comparison of the maximum posterior value with uncertainty quoted at the 90\% level and the credible level at which the true value is recovered ($\mathrm{CL_{inj}}$) for the component mass and spin parameters using both the mass and spin sorting for the simulated signal. The credible level is calculated using the highest posterior density  method.
}
\label{tab:injection}
\begin{ruledtabular}
\begin{tabular}{l  l r r r r}
	Parameter & Inj.  & \multicolumn{2}{c}{Mass sorting} & \multicolumn{2}{c}{Spin sorting}\\
     \cline{3-4}
     \cline{5-6}
     &  & maxP & $\mathrm{CL}_{\mathrm{inj}}$ & maxP & $\mathrm{CL}_{\mathrm{inj}}$ \\
    \midrule
    $m_{1/A}$      & $40~\mathrm{M}_{\odot}$ & $40.90^{+3.02}_{-1.43}$ & 58.5\% & $39.27^{+3.77}_{-2.88}$ & 45.7\% \\[3pt]
    $m_{2/B}$      & $40~\mathrm{M}_{\odot}$ & $38.70^{+1.74}_{-2.38}$ & 67.2\% & $40.19^{+3.05}_{-3.05}$ & 0\% \\[3pt]
    $\chi_{1/A}$   & 0.8 & $0.01^{+0.85}_{-0.01}$ & 87.1\% & $0.77^{+0.21}_{-0.17}$ & 11.5\% \\[3pt]
    $\chi_{2/B}$   & 0 & $0.80^{+0.19}_{-0.80}$ & 61.4\% & $0.01^{+0.41}_{-0.01}$ & 0\% \\[3pt]
    $\theta_{1/A}$ & 1.57 rad & $1.54^{+0.87}_{-0.80}$ & 0\% & $1.59^{+0.29}_{-0.34}$ & 0\% \\[3pt]
    $\theta_{2/B}$ & --  & $1.62^{+0.73}_{-0.58}$ & -- & $1.54^{+1.20}_{-0.81}$ & -- \\
\end{tabular}
\end{ruledtabular}
\end{table}

\begin{figure*}
	\centering
	\includegraphics[width=0.8\columnwidth]{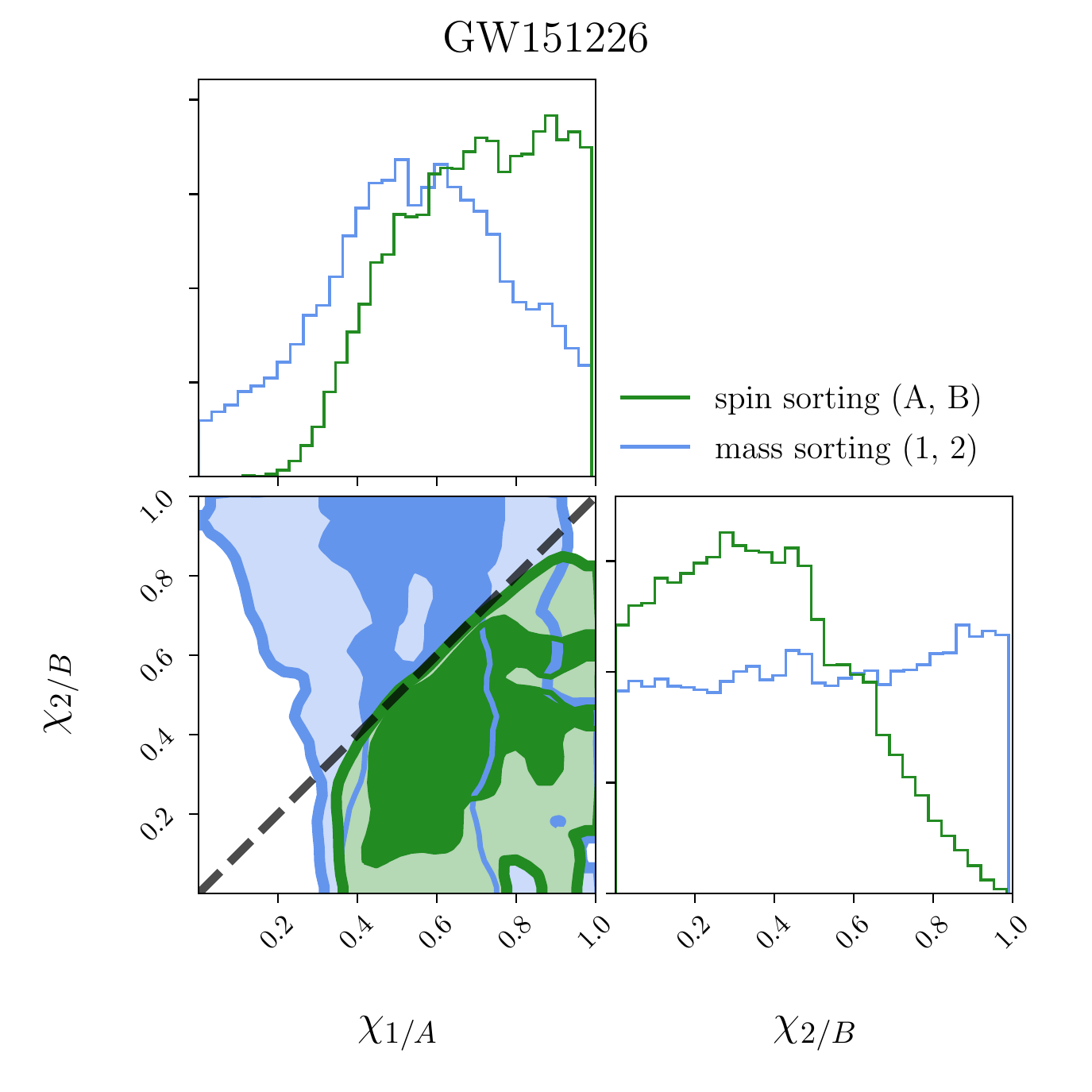}\qquad
	\includegraphics[width=0.8\columnwidth]{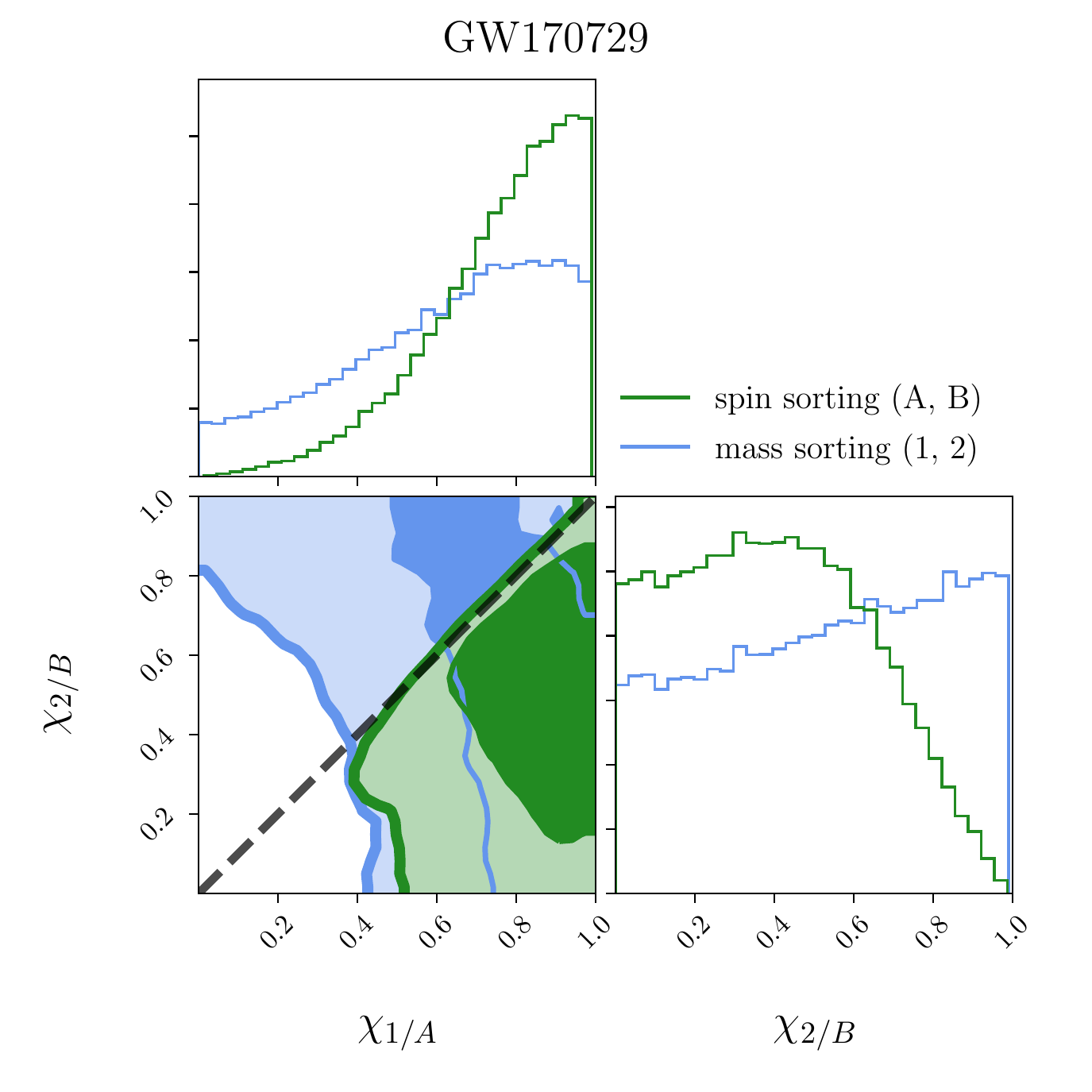}
	\caption{Comparison corner plot for the spin magnitudes for the posteriors obtained using the IMRPhenomPv2 waveform for GW151226 and GW170729. The equal-spin diagonal is shown as a dashed line for reference.}
	\label{fig:GW151226}
\end{figure*}

In Fig.~\ref{fig:injection}, we compare the resulting measurements of the spin magnitudes and tilts using both the mass and spin sortings.
The mass sorting induces a bimodal posterior in the $\chi_{1}, \chi_{2}$ plane (symmetric around $\chi_1=\chi_2$), showing that a high spin could be assigned to either component, while the marginal posteriors on $\chi_{1}$ and $\chi_{2}$ are largely unconstrained.
The spin sorting breaks this degeneracy, restricting the posterior so that $\chi_{A}$ peaks at the true value and $\chi_{B}$ rails against the lower edge of the prior.
A similar degeneracy can be seen in the two-dimensional posterior for $\theta_{1}$ and $\theta_{2}$ in the mass sorting, which shows that information is retrieved for the tilt of one of the two objects without identifying which.
Switching to the spin sorting, the $\theta_{A}$ posterior is well-constrained, while $\theta_{B}$ returns the prior.
Thus, this parametrization makes it clear that we can measure the tilt of the highest-spin object but cannot say anything about the lowest-spin one, as expected given that $\chi_B=0$, making $\theta_{B}$ irrelevant.
In Table~\ref{tab:injection}, we show the maximum posterior probability values and associated 90\% credible interval for the component masses, spins, and tilt angles obtained using both parameterizations. We also show the credible level at which the true value is recovered.
The mass ratio posterior is peaked narrowly around the equal-mass limit, $q=0.996^{+0.004}_{-0.138}$.
The statistical uncertainty for the component masses $m_{A}$ and $m_{B}$ is greater than for $m_{1}$ and $m_{2}$ because there is no imposed ordering on $m_{A}$ and $m_{B}$.

In order to verify that the improved resolution of the spin sorting is robust against changes in the true spin magnitude and tilt and that it extends to systems without exactly equal component masses, we repeat our simulation for a variety of different binary parameters.
We find a similar improvement in the resolution of the component spins for systems with lower spins, $\chi_{A} = 0.2$, with an aligned primary spin, $\theta_{A} = 0$, and with slightly unequal mass ratios, $q= m_{2}/m_{1} = 0.9$, when each of these parameters is varied independently in simulated signals with the same SNR as the original.
When the network SNR is decreased to 12,
the component spins cannot be well-measured using either parameterization, meaning that only minor deviations from the priors are observed.

The spin sorting ceases to be useful for systems where the mass ratio is measurably different from unity. Looking at a system with $q=0.7$ and ${\rm SNR}=30$, the spin sorting introduces the same type of degeneracies in the spin parameters as are present in Fig.~\ref{fig:injection} under the mass sorting.
This is because the spin of the most massive object is well-defined for systems where the most massive object can be distinguished.
Systems with equal spin magnitudes---both nonspinning and highly spinning with $\chi_{1}=\chi_{2}=0.8$---similarly do not benefit from the spin sorting, as the spin magnitude posteriors are largely unchanged in this case. (The $\chi_{A}$ posterior for the nonspinning injection peaks at $\chi_{A}=0$, even though this region is disfavored by the prior.) However, when analyzing a system with $\chi_{1}=\chi_{2}=0.8$ and unequal tilt angles, $\theta_{1}=\pi/2$ and $\theta_{2}=0$, the bimodality in the tilt posterior can be resolved by instead sorting by the tilt angles without affecting the spin magnitude posteriors.

\heading{LIGO-Virgo detections}
We apply the same reparameterization to the publicly released posterior samples for the first 13 LIGO-Virgo detections~\cite{Vallisneri:2014vxa, Abbott:2019ebz, LIGOScientific:2018mvr, Abbott:2020uma, LIGOScientific:2020stg}.
The ten BBH mergers announced in the first LIGO-Virgo catalog (GWTC-1) are all consistent with $q=1$, although the posteriors all support considerably lower values than the simulated signal in Fig.~\ref{fig:injection}.
For these systems, we find that the differences between the spin and mass sorting are generally not as significant as for the simulation.
This is consistent with the results of the low-SNR simulation discussed above.

For the two events whose posteriors in the mass sorting already indicated a preference for non-zero spins, GW151226~\cite{Abbott:2016nmj} and GW170729~\cite{Chatziioannou:2019dsz}, $\chi_{A}=0$ is ruled out with $3\sigma$ credibility. For GW170729, $\chi_A=\chi_B=1$ is included within the 90\% credible region, while for GW151226, $\chi_{A}=1$ is included in the 50\% credible region as long as $0.5< \chi_{B} < 0.7$.
We show spin magnitude posteriors for these events using both parameterizations in Fig.~\ref{fig:GW151226}.
For all the BBH systems analyzed, the lower bound of the 90\% credible interval for the $\chi_{A}$ posterior is $\chi_{A}\geq 0.14$, although this is dominated by the triangular prior (grey line in Fig.~\ref{fig:injection}). On the other hand, the $\chi_{1}$ posterior is only constrained to $\chi_{1}>0.14$ for GW151226 and GW170729 under the mass sorting.
For the unequal-mass binary GW190412~\cite{LIGOScientific:2020stg},
the spin sorting introduces degeneracies in the spin parameters that were not present in the mass-based sorting, which we expected from our $q<1$ simulations.
In the Supplementary Material, we explore the features of the posteriors for GW190412 and the two binary NS systems and show that waveform systematics proved to be important for some events (including GW150914).

\heading{Population analyses}
In order to determine the effects of the spin sorting on the inferred population properties of BH spins, we use the infrastructure of hierarchical Bayesian inference to characterize the underlying distributions of $\chi_{A}$ and $\chi_{B}$. If the mass-sorted spin magnitudes for individual events, $\chi_{1/2}$, are modeled as being drawn from the same Beta distribution following \cite{Wysocki:2018, LIGOScientific:2018jsj}, we can compute the corresponding $\chi_{A}$ and $\chi_{B}$ distributions using order statistics, by assuming they correspond to the maximum and minimum of two draws from the $\chi_{1/2}$ distribution. We use the publicly released posterior samples for the  $\chi_{1/2}$ hyperparameters from LVC analyses first including only GWTC-1 events~\cite{LIGOScientific:2018jsj, data_release_O2}, then including all 44 confident BBH detections reported in GWTC-2~\cite{Abbott:2020niy, Abbott:2020gyp, data_release_O3}. We present our results using the posterior population distribution (PPD), which is the expected distribution for the individual-event parameters of new BBH events inferred from the accumulated set of detections (e.g., \cite{LIGOScientific:2018jsj}, see Supplementary Material).

In Figure~\ref{fig:iid_chiI_chi_II} we show the PPDs for $p(\chi_{A})$ and $p(\chi_{B})$ as well as the 50\% and 90\% credible bands. 
The inferred distribution for $\chi_{A}$ (top) peaks at around $\chi_{A} \sim 0.3$ and has negligible support for $p(\chi_{A}) = 0$ for both the GWTC-1 and GWTC-2 analyses, indicating that most of the highest-spinning BHs in LIGO-Virgo binaries have nonzero spins. This is very different from the distribution inferred for the mass sorted spins, $p(\chi_{1/2})$, which has considerable support at $\chi_{1/2}=0$ (see Fig. 8 of \cite{LIGOScientific:2018jsj} and Fig. 10 of \cite{Abbott:2020gyp}).
The lower bound of the 50\% credible interval for $p(\chi_{A})$ is $\chi_{A}=0.19$, while the 50\% credible interval for $\chi_{1/2}$ extends down to $\chi_{1/2}=0.06$ for the GWTC-2 analysis. 
The distribution of spin magnitudes for the lowest-spinning BHs (bottom) is consistent with peaking at $\chi_{B}=0$ for both analyses and has more posterior support at $\chi_{B}=0$ when including the full GWTC-2 sample.
All these distributions vary significantly from those obtained using samples from the spin magnitude prior instead of posterior samples for the 44 confident BBH detections from GWTC-2 (dashed lines in Fig.~\ref{fig:iid_chiI_chi_II}).



\begin{figure}[h!]
	\centering
	\includegraphics[width=\linewidth]{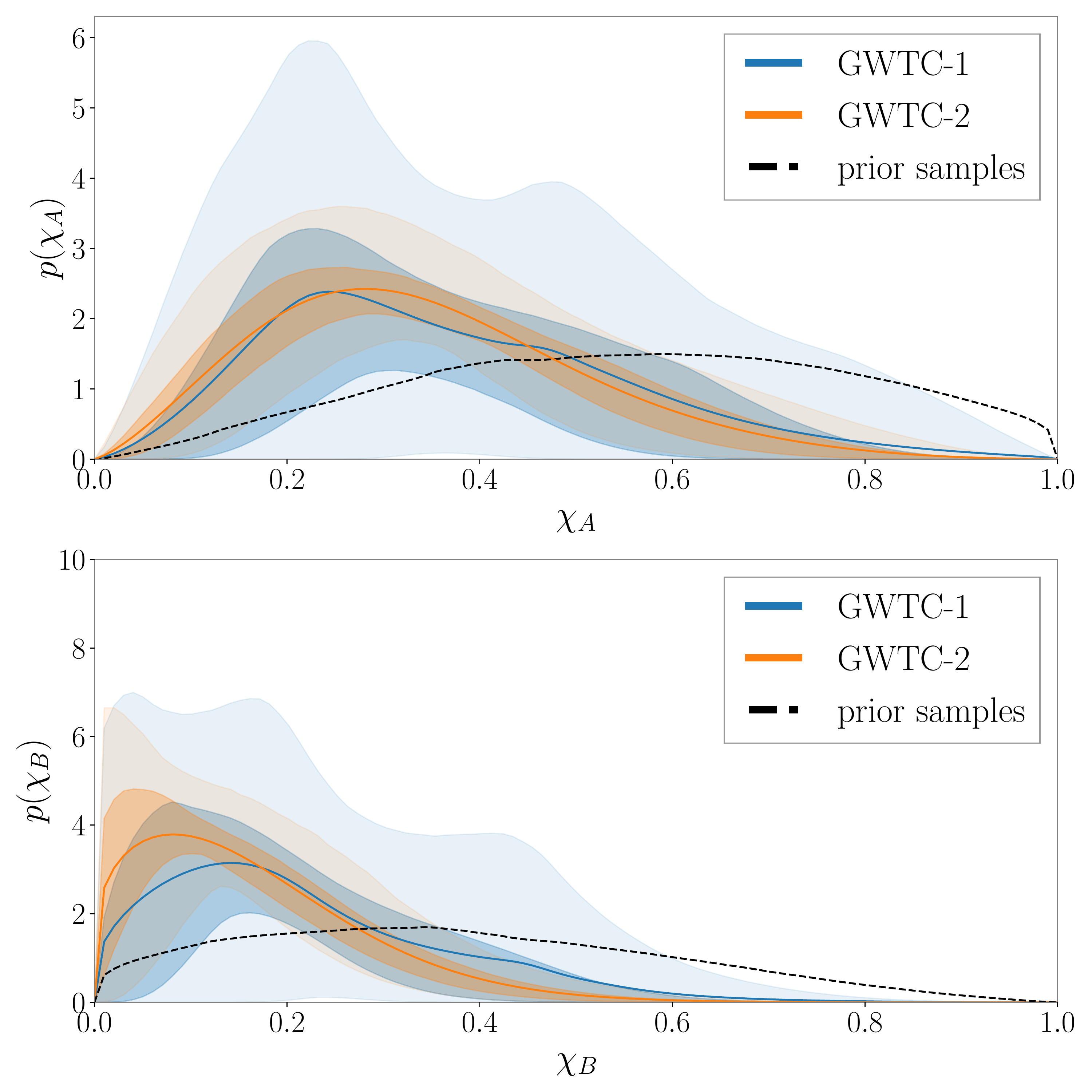}
	\caption{PPDs for $p(\chi_{A})$ (top) and $p(\chi_{B})$ (bottom). The blue curves include only events from GWTC-1, while the orange curves include all 44 confident BBH events in GWTC-2. The shaded regions correspond to the 50\% and 90\% credible intervals, and the PPDs obtained using prior samples for the individual GWTC-2 events are shown in the dashed lines.}
	\label{fig:iid_chiI_chi_II}
\end{figure}

\begin{figure}
	\centering
	\includegraphics[width=\linewidth]{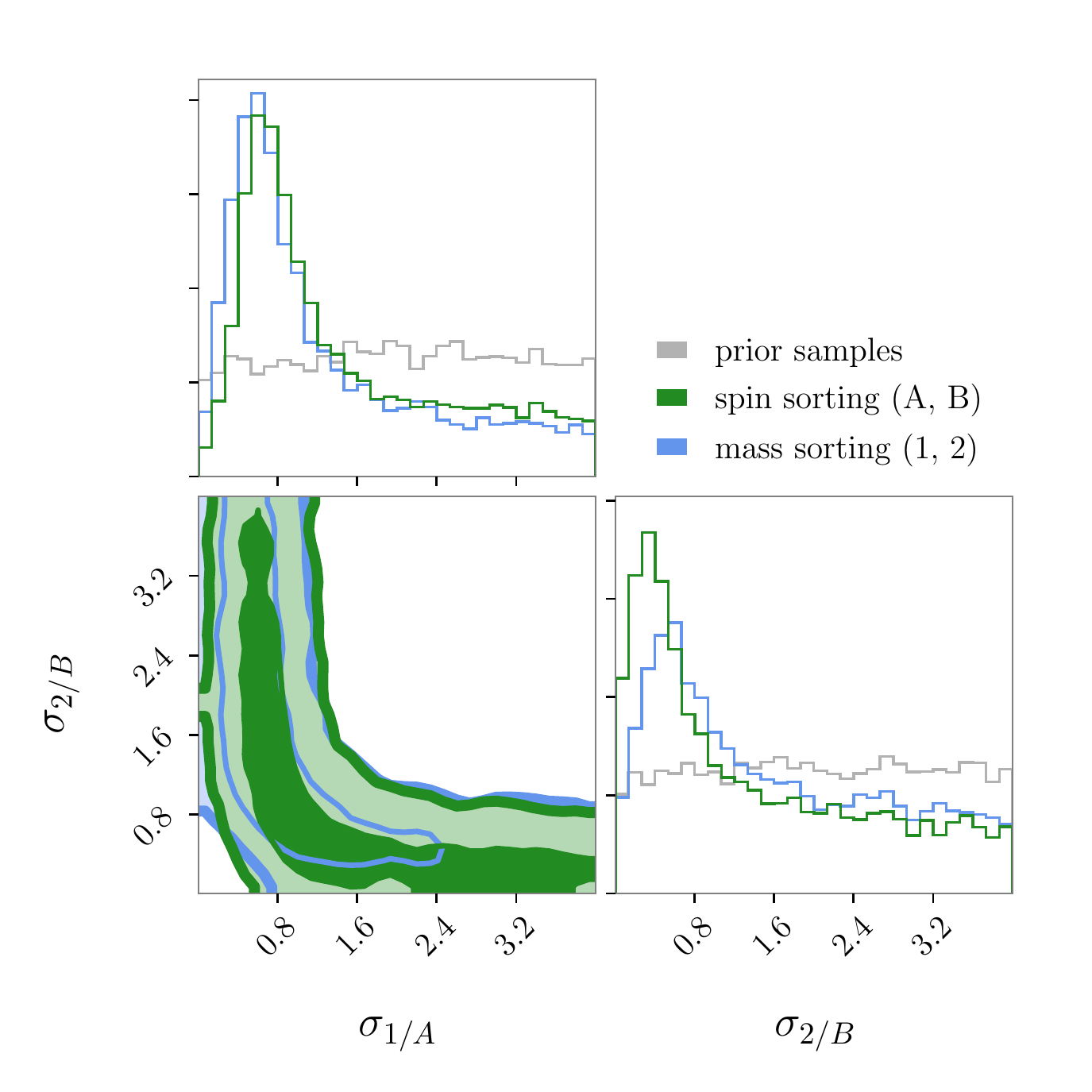}
	\caption{Corner plot comparing the inference on the spin tilt hyperparameters using the mass sorted tilts $\theta_{1}$ and $\theta_{2}$ to the posteriors obtained using the same population model but for the spin sorted tilts $\theta_{A}$ and $\theta_{B}$ for the GWTC-2 events. The posteriors obtained using prior samples for the individual events are shown in grey.}
	\label{fig:tilt_corner_mixture}
\end{figure}

For the spin tilt angles, we follow \cite{Talbot:2017yur, LIGOScientific:2018jsj} and model the distribution as the sum of two populations motivated by the most popular BBH formation channels (eg.~\cite{Kalogera:1999tq, Schnittman:2004, Postnov:2014tza, Rodriguez:2016vmx}): an isotropic component and a preferentially aligned component, where the hyperparameters $\sigma_{1/A}$ and $\sigma_{2/B}$ control the spread in the possible tilt angles around $\theta_{1/A}= \theta_{2/B} = 0$. A nonzero value for $\sigma$ indicates that not all tilts are aligned. We conduct hierarchical Bayesian inference using this model for the tilt angles under both the mass and spin sorting. 
In Fig.~\ref{fig:tilt_corner_mixture}, we show the posteriors on the hyperparameters describing the spin tilt population model for the 44 confident BBH detections in GWTC-2.
The blue distribution corresponds to inference starting from the mass-sorted single-event posteriors, while the green distribution shows the same for the spin-sorted posteriors. 
The posterior for $\sigma_{A}$ is constrained slightly further away from 0 than that of $\sigma_{1}$, while the opposite is true for $\sigma_{B}$ and $\sigma_{2}$. This indicates that the posterior information gain relative to the prior is less balanced between the two tilt angles in the spin sorting than the mass sorting.

We highlight that $\sigma_{1/A} = \sigma_{2/B} = 0$ is excluded with $>99\%$ credibility for both parameterizations when marginalized over spin magnitude. The same is true at 90\% credibility when considering only the GWTC-1 events.
This means that the preferentially aligned component is more likely to have nonzero width, and hence a fraction of binaries is likely to have in-plane spin components. 
This result agrees with previous analyses, which find that a fully-aligned population is disfavored by the LVC detections reported in GWTC-1~\cite{Tiwari:2018qch, LIGOScientific:2018jsj, Wysocki_2019}, and that the current detections are consistent with a population of high spins that are significantly misaligned with respect to the orbital angular momentum.~\cite{Farr:2017uvj, Miller:2020zox}. 
The GWTC-2 results agree with those presented in \cite{Abbott:2020gyp} using two different spin tilt parameterizations from the one we use, which find evidence for general-relativistic spin precession at the population level.
We confirm that this feature originates in the data---and is not just an artifact of the Monte Carlo integration performed during hierarchical inference step---by replacing the mass-sorted spin tilt posteriors from individual events with draws from the prior. This results in an uninformative distribution for $\sigma_{1}, \sigma_{2}$ consistent with having uniform support across the prior range, including the region around $\sigma_{1} = \sigma_{2} = 0$ (shown in grey in Fig.~\ref{fig:tilt_corner_mixture}). The region $\sigma_{1/A}, \sigma_{2/B} \geq 1.6$ is also excluded at $> 90\%$ credibility, indicating that a fully isotropic spin distribution corresponding to high values of $\sigma$ is statistically disfavored. Additional analysis details are provided in the Supplementary Material.

\heading{Conclusion}
We have demonstrated the advantages of introducing an alternative labeling for the component objects of a compact binary based on the spins instead of the masses; we denote the object with the largest (smallest) spin magnitude by A (B), such that $\chi_{A} > \chi_{B}$. Through analysis of simulated signals, we find that this sorting improves the resolution of the component spins of binaries consistent with having equal mass, regardless of the magnitude and tilt of the primary spin. When applied to the posteriors for the GWTC-1 events, we find that the most-spinning object is consistent with having extremal spin for the events that were already known to prefer nonzero spins, GW151226 and GW170729. 
The spin sorting ceases to be useful for systems with measurably unequal masses, as was the case for GW190412.

We characterize the distributions of $\chi_{A}, \chi_{B}, \theta_{A}, \theta_{B}$ of the 44 confident GWTC-2 BBH events by means of hierarchical Bayesian inference. Modeling the mass-sorted spins as drawn from a Beta distribution, we compute the implied probability densities for $\chi_{A}$ and $\chi_{B}$ and find that $p(\chi_{A})$ peaks at $\chi_A \sim 0.3$ and has negligible support at $\chi_{A}=0$, while $p(\chi_{B})$ and $p(\chi_{1/2})$ have considerable posterior support at $\chi=0$. We thus conclude, at the population level, that most of the BBHs detected by LIGO-Virgo have at least one component with nonzero spin.
When modeling the distributions of spin tilt angles, we find that the configuration where all of the spins in the population are aligned with the orbital angular momentum is excluded from the 99\% credible interval for both the mass and the spin sorting.

We end by stressing that the spin sorting does not introduce new information into the analysis. We are not applying different priors, but rather defining a new set of parameters that can be inferred using the same individual-event posterior samples used in the original mass-sorting. This implies that the Bayesian evidence of the data is unchanged.
The reparameterization can be done entirely in post-processing and does not affect the posteriors for parameters that do not distinguish between the binary components like the effective aligned and precessing spins, $\chi_{\rm eff}$ and $\chi_{\rm p}$.
The spin sorting also does not change existing population-level inferences obtained from the mass-sorted component parameters, with the possible exception of analyses relying on marginalized one-dimensional posteriors on the component spin quantities, which cannot encode the parameter degeneracies that we noted for signals consistent with $q\approx 1$.
While in some cases BBH formation channels make it more natural to label the component objects based on their mass, thinking about the objects primarily in terms of spin could lead to rich new ways to test astrophysical models moving forward~\footnote{The Supplementary Material contains more details on the Bayesian methods employed for both the individual-event analyses and the hierarchical inference, as well as additional information for the features of the posteriors of GW190412, the two binary NS systems, and an exploration of the effect of waveform systematics. This further includes Refs. \cite{Romano:2016dpx, Varma:2019csw, Talbot:2018cva, Loredo_2004, Fishbach_2018, Thrane:2018qnx, Mandel_2019, Vitale:2020aaz, data_release_VT, Farr_2019, data_release_samples, Speagle_2020, Talbot:2019, Husa:2015iqa, Hannam:2013oca, Khan:2015jqa, Babak:2016tgq,Taracchini:2013rva, Pan:2013rra, TheLIGOScientific:2017qsa, TheLIGOScientific:2016wfe, Hoy:2020vys}}.


%


\begin{acknowledgments}
\heading{Acknowledgments}
S.B., M.I. and S.V.\ acknowledge support of the National Science Foundation and the LIGO Laboratory.
LIGO was constructed by the California Institute of Technology and
Massachusetts Institute of Technology with funding from the National
Science Foundation and operates under cooperative agreement PHY-0757058.
S.B. is also supported by the Paul and Daisy Soros Fellowship for New Americans and the NSF Graduate Research Fellowship under Grant No. DGE-1122374.
M.I.\ is supported by NASA through the NASA Hubble Fellowship
grant No.\ HST-HF2-51410.001-A awarded by the Space Telescope
Science Institute, which is operated by the Association of Universities
for Research in Astronomy, Inc., for NASA, under contract NAS5-26555.
V.V.\ is generously supported by the Sherman Fairchild Foundation, and NSF
grants PHY–170212 and PHY–1708213 at Caltech, and by a Klarman Fellowship at
Cornell.
The authors would like to thank Colm Talbot for help with \texttt{GWPopulation}, and Thomas Callister, Thomas Dent, Maya Fishbach, and Eric Thrane for careful comments on the manuscript. We also thank Will Farr, Katerina Chatziioannou, Javier Roulet, Leo Stein, Simona Miller, Carl-Johan Haster, Saavik Ford, Barry McKernan, Daniel Wysocki, Richard O'Shaughnessy, and others for useful discussions.
This paper carries LIGO document number LIGO-P2000247.
\end{acknowledgments}

\bibliography{spins}
\clearpage
\section*{Supplementary material}
\label{supp_mat}
\setcounter{equation}{0}
\setcounter{figure}{0}
\setcounter{table}{0}

\subsection{Parameter estimation methods}
To analyze the simulated signals, we perform Bayesian parameter estimation using the standard Gaussian likelihood for gravitational-wave data~\cite{Romano:2016dpx, Veitch:2014wba}:
\begin{align}
\mathcal{L}(d_{i} |\theta) = \prod_{j} \frac{2}{\pi T S_n(f_{j})}\exp{\left[-\frac{2|d_{i}(f_{j}) - h(f_{j};\theta)|^{2}}{T S_n(f_{j})}\right]},
\label{eq:likelihood}
\end{align}
where $T$ is the duration of the data segment being analyzed, $S_n(f_{j})$ is the noise power spectral density of the detector, $d_{i}(f_{j})$ is the strain data for event $i$, and $h(f_{j};\theta)$ is the waveform model for the compact binary source. We simulate the measurement using the \texttt{LALInference} algorithm~\cite{Veitch:2014wba} and the numerical relativity surrogate waveform model NRSur7dq4~\cite{Varma:2019csw}. In order to obtain posterior samples for the parameters $\theta$ using the likelihood in Eq.~\ref{eq:likelihood}, we impose priors that are uniform in the component masses $m_{1},\ m_{2}$ between $10~\mathrm{M_{\odot}}$ and $240~\mathrm{M_{\odot}}$, with constraints on the total mass between $70~\mathrm{M_{\odot}}$ and $240~\mathrm{M_{\odot}}$ and on mass ratio $q$ between 0.2 and 1. The luminosity distance prior is $\propto d_{L}^{2}$ over the range 1--7000~Mpc.

For the spin tilt population inference, we simultaneously fit the mass and spin magnitude distributions. We follow \cite{Wysocki:2018, LIGOScientific:2018jsj} and assume that both of the black hole (BH) spin magnitudes under the mass sorting are drawn from the same Beta distribution with hyperparameters $\alpha$ and $\beta$,
\begin{align}
p(\chi_{1/2} | \alpha, \beta) = \frac{\chi_{1/2}^{\alpha -1}(1-\chi_{1/2})^{\beta-1}}{B(\alpha, \beta)},
\label{eq:beta}
\end{align}
where $B(\alpha, \beta)$ is the Beta function. We restrict the priors on the Beta function parameters to exclude values of $\alpha, \beta \leq 1$ corresponding to singular Beta distributions. This means that $p(\chi)$ \emph{must} peak within $0 < \chi < 1$, as nonsingular Beta distributions vanish at those values. The Beta distribution described in Eq.~\ref{eq:beta} can also be parameterized in terms of its mean and variance:
\begin{align}
\mu(\chi) &= \frac{\alpha}{\alpha + \beta},\\
\sigma^{2}(\chi) &= \frac{\alpha \beta}{(\alpha + \beta)^{2}(\alpha + \beta + 1)}.
\end{align}
We choose to sample in $\mu(\chi)$ and $\sigma^{2}(\chi)$, imposing constraints such that $\alpha, \beta > 1$ to restrict the parameter space to only nonsingular Beta distributions.

The distributions for $\chi_{A}$ and $\chi_{B}$ obtained by applying order statistics to the Beta distribution for $\chi_{1/2}$ are given by:
\begin{align}
p(\chi_{A}) &= 2\, p(\chi_{1/2} | \alpha, \beta)\, \mathrm{CDF}(\chi_{1/2} | \alpha, \beta)\, ,\\
p(\chi_{B}) &= 2\, p(\chi_{1/2} | \alpha, \beta)\left[1-\mathrm{CDF}(\chi_{1/2}| \alpha, \beta)\right],
\end{align}
assuming $\chi_{A}$ is the maximum and $\chi_{B}$ is the minimum of two draws from $p(\chi_{1/2}| \alpha, \beta)$.
$\mathrm{CDF}(\chi_{1/2})$ is the cumulative distribution function for $\chi_{1/2}$ given by the regularized incomplete Beta function with parameters $(\alpha, \beta)$.

We further assume that the primary mass distribution is described by the sum of a truncated power-law with low-mass smoothing and a Gaussian component~\cite{Talbot:2018cva}. The hyperparameters describing this model are the slope $\alpha_{m}$, upper and lower cutoffs $m_{\max}$ and $m_{\min}$, the low-mass smoothing parameter $\delta_{m}$, the peak and width of the Gaussian component $\mu_{m}$ and $\sigma_{m}$, and the mixing fraction between the two components $\lambda_{\mathrm{peak}}$.
The mass ratio distribution is modeled as a power law with slope $\beta_{q}$. 
This corresponds to Model C from \cite{LIGOScientific:2018jsj}, dubbed``power-law + peak''.
We use this same mass model when computing the spin magnitude distribution using prior samples for the GWTC-2 events, shown in the dashed black line in Fig. 3 in the main text.

Following \cite{Talbot:2017yur}, the distribution for spin tilt angles is given by the sum of an isotropic component and a preferentially aligned component, which is composed of the product of two truncated Gaussians peaked at $\cos{t_{i}}=1$ for each tilt angle:
\begin{align}
\label{eq:tilts}
p(\cos{t_{1/A}}, \cos{t_{2/B}} |\sigma_{1/A}, \sigma_{2/B}, \xi) = \frac{1-\xi}{4} + \\ \nonumber
\frac{2\xi}{\pi}\prod_{i \in \{1/A, 2/B\}} \frac{\exp(-(1-\cos{t_{i}})^{2}/(2\sigma_{i}^{2}))}{\sigma_{i}\mathrm{erf}(\sqrt{2}/\sigma_{i})},
\end{align}
where the hyperparameter $\xi$ gives the mixture fraction between the two components. The complete population model, $\pi(\theta | \Lambda)$, is given by the product of Eqs.~\ref{eq:beta}, \ref{eq:tilts}, and the ``power-law + peak'' mass distribution.

Using the population model $\pi(\theta | \Lambda)$ to describe the distribution of individual-event parameters $\theta$, the likelihood of observing the hierarchical parameters $\Lambda$ for a data set $\{d\}$ consisting of $N_{\mathrm{det}}$ detected events is given by:
\begin{align}
\mathcal{L}(\{d\} | \Lambda) \propto \prod_{i=1}^{N_{\mathrm{det}}} \frac{\int \mathcal{L}(d_{i} | \theta)\pi(\theta | \Lambda)}{\alpha(\Lambda)}
\label{eq:hyper-likelihood}
\end{align}
where $\alpha(\Lambda)$ represents the detectable fraction of events assuming the individual-event parameters are drawn from distributions specified by hyperparameters $\Lambda$~\cite{Loredo_2004, Fishbach_2018, Thrane:2018qnx, Wysocki_2019, Mandel_2019, Vitale:2020aaz}. We use the sensitive spacetime volume estimates released by the LVC in \cite{data_release_VT}, which for the GWTC-2 analysis were determined through injection campaign~\cite{Abbott:2020niy} and for the GWTC-1 events were obtained using simulated data. We calculate $\alpha(\Lambda)$ using the formalism described in \cite{Farr_2019}. We do not account for the selection biases due to the spin parameters, since those have a much smaller effect than the mass parameters~\cite{Abbott:2020gyp}.

The likelihood in Eq.~\ref{eq:hyper-likelihood} is evaluated using a Monte Carlo integral over the individual-event parameter posteriors released by the LVC for the binary BH (BBH) events included in GWTC-1~\cite{LIGOScientific:2018mvr, Abbott:2019ebz} and GWTC-2~\cite{Abbott:2020niy, data_release_samples}. For the GWTC-1 events, we use the samples obtained with the IMRPhenomPv2 waveform model, while for GWTC-2 we use the ``Publication'' posterior samples presented in ~\cite{Abbott:2020niy}, which in most cases use a combination of waveform models including the effects of spin precession and higher-order multipoles. For GWTC-2, we only analyze the 44 confident BBH detections with a false alarm rate $< 1~\mathrm{yr}^{-1}$ and exclude the events where at least one of the compact objects has considerable posterior support below $3~\mathrm{M}_{\odot}$.
We use the \texttt{dynesty}~\cite{Speagle_2020} sampler, as implemented in the \texttt{GWPopulation}~\cite{Talbot:2019} package, to obtain hyperparameter posterior samples. 

The posterior population distribution calculated using the hyperparameter posteriors obtained using the likelihood in Eq.~\ref{eq:hyper-likelihood} is given by:
\begin{align}
\mathrm{PPD}(\theta | \{d\}) = \int \pi(\theta | \Lambda)p(\Lambda | \{d\})d\Lambda.
\end{align}

\begin{table*}
    \centering
\begin{tabular}{|p{1.5cm} ||p{3cm} p{2.5cm} p{2.5cm}|}
    \hline
    Parameter & Prior & GWTC-1 & GWTC-2\\
    \hline
    $\alpha_{m}$ & U(-4, 12) & $6.41^{+4.87}_{-4.44}$ & $2.96^{+0.77}_{-0.63}$\\
    $\beta_{m}$  & U(-4, 12) & $5.60^{+5.65}_{-5.73}$  & $1.01^{+2.26}_{-1.41}$ \\
    $m_{\max}$ & $\mathrm{U}(30~\mathrm{M}_{\odot}, 100~\mathrm{M}_{\odot})$ & $59.80^{+35.92}_{-26.66}~\mathrm{M}_{\odot}$ & $86.30^{+12.10}_{-12.66}~\mathrm{M}_{\odot}$\\
    $m_{\min}$  & $\mathrm{U}(2~\mathrm{M}_{\odot}, 10~\mathrm{M}_{\odot})$ & $7.40^{+1.35}_{-3.31}~\mathrm{M}_{\odot}$ & $4.71^{+1.38}_{-1.84}~\mathrm{M}_{\odot}$\\
    $\delta_{m}$  & $\mathrm{U}(0~\mathrm{M}_{\odot}, 10~\mathrm{M}_{\odot})$ & $2.58^{+5.28}_{-2.38}~\mathrm{M}_{\odot}$ & $4.59^{+4.24.35}_{-3.98}~\mathrm{M}_{\odot}$\\
    $\mu_{m}$ & $\mathrm{U}(20~\mathrm{M}_{\odot}, 50~\mathrm{M}_{\odot})$ & $28.60^{+5.71}_{-6.98}~\mathrm{M}_{\odot}$ & $32.42^{+3.52}_{-5.91}~\mathrm{M}_{\odot}$ \\
    $\sigma_{m}$ & $\mathrm{U}(0.4~\mathrm{M}_{\odot}, 10~\mathrm{M}_{\odot})$ & $6.13^{+3.37}_{-4.18}~\mathrm{M}_{\odot}$ & $5.17^{+4.21}_{-3.85}~\mathrm{M}_{\odot}$ \\
    $\lambda_{\mathrm{peak}}$ & $\mathrm{U}(0, 1)$ & $0.19^{+0.40}_{-0.16}$ & $0.07^{+0.13}_{-0.05}$ \\
    $\mu(\chi)$ & $\mathrm{U}(0, 1)$ & $0.30^{+0.20}_{-0.15}$ & $0.31^{+0.11}_{-0.09}$ \\
    $\sigma^{2}(\chi)$ & $\mathrm{U}(0, 0.25)$ & $0.02^{+0.03}_{-0.02}$ & $0.03^{+0.02}_{-0.02}$ \\
    $\xi$ & $\mathrm{U}(0, 1)$ & $0.54^{+0.41}_{-0.47}$ & $0.79^{+0.19}_{-0.43}$ \\
    \hline
\end{tabular}
\caption{Priors, posterior medians, and 90\% credible intervals for the hyperparameters used in our population analysis obtained for spin-sorted component tilt angles including both GWTC-1 events alone and the 44 confident BBH detections in GWTC-2. All the priors are uniform across the specified range and match those used in ~\cite{Abbott:2020gyp}.}
\label{tab:hyper_param_priors}
\end{table*}

The priors and posterior results for all hyperparameters except $\sigma_{1/A}$ and $\sigma_{2/B}$ for the spin-sorted tilt inference performed using the GWTC-2 BBH events are shown in Table~\ref{tab:hyper_param_priors}. The priors are uniform for all parameters and identical for both the mass and spin-sorted tilt analysis. We use uniform priors over the range $(0,4)$ for the tilt $\sigma$ parameters. The posterior for the spin tilt mixing fraction $\xi$ peaks at the upper edge of its prior, indicating that a purely isotropic distribution of tilts is statistically disfavored by the data. We obtain posteriors for the mass, spin magnitude, and $\xi$ hyperparameters comparable to those quoted in \cite{Abbott:2020gyp}.  The results for these parameters change negligibly under the mass and spin sortings.

\begin{figure}
	\centering
	\includegraphics[width=\linewidth]{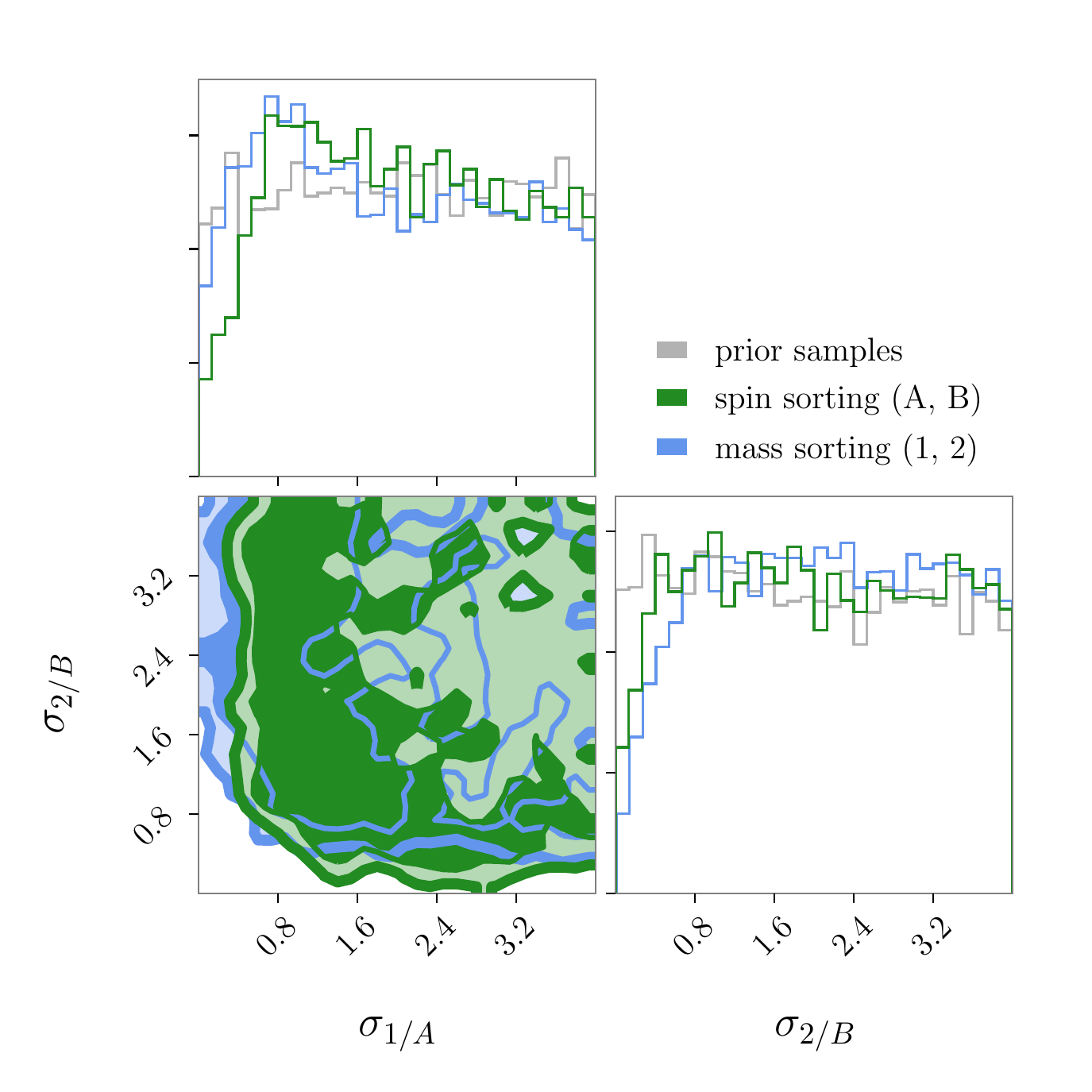}
	\caption{Corner plot comparing the inference on the spin tilt hyperparameters using the mass sorted tilts $\theta_{1}$ and $\theta_{2}$ and spin sorted tilts $\theta_{A}$ and $\theta_{B}$ for the GWTC-1 events. The posteriors obtained using prior samples for the individual events are shown in grey.}
	\label{fig:tilts}
\end{figure}

The corner plot of the tilt $\sigma$ hyperparameters for the GWTC-1 analysis is shown in Fig.~\ref{fig:tilts}. Even with only the GWTC-1 events, the corner of parameter space at $\sigma_{1/A} = \sigma_{2/B} = 0$ representing a fully-aligned population is excluded at $>90\%$ credibility for both the mass and spin sortings.  The posterior for $\sigma_{B}$ is less constrained that that for $\sigma_{2}$---indicating that for this analysis, the information on the tilt angles at the population level is predominantly obtained from the measurement of the highest spinning object, rather than the most massive. A similar trend is present in the $\sigma$ posteriors for the full GWTC-2 analysis shown in Fig. 4 in the main text.

\subsection{Additional results for individual sources}
\begin{figure}
	\centering
	\includegraphics[width=0.99\linewidth]{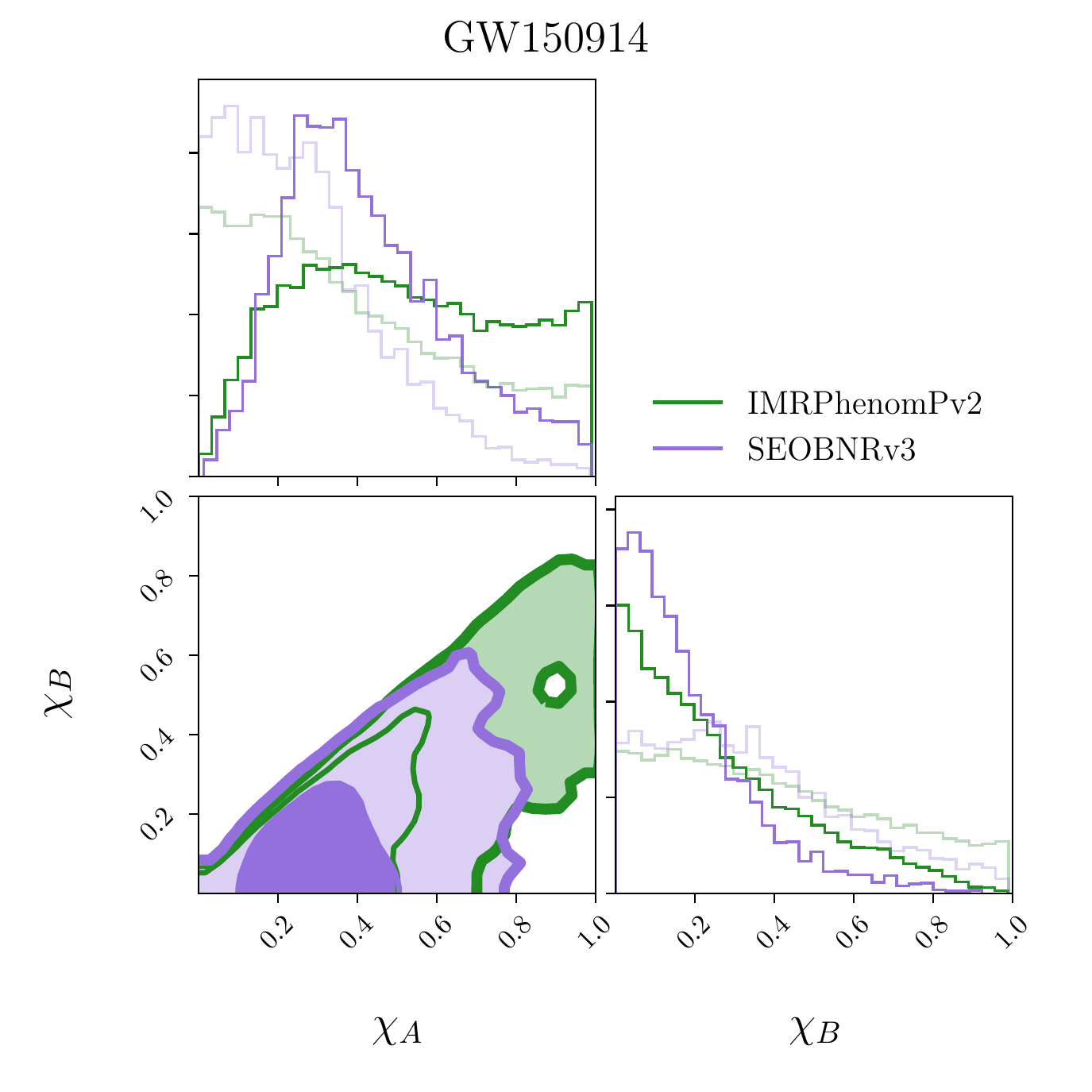}
	\caption{Comparison corner plot for the spin magnitudes for the posteriors obtained using both the IMRPhenomPv2 (green) and SEOBNRv3 (purple) waveforms for GW150914 using the spin sorting. The marginalized posteriors obtained using the mass sorting are shown in lighter green and purple.}
	\label{fig:GW150914}
\end{figure}

\begin{figure}
	\centering
	\includegraphics[width=0.99\linewidth]{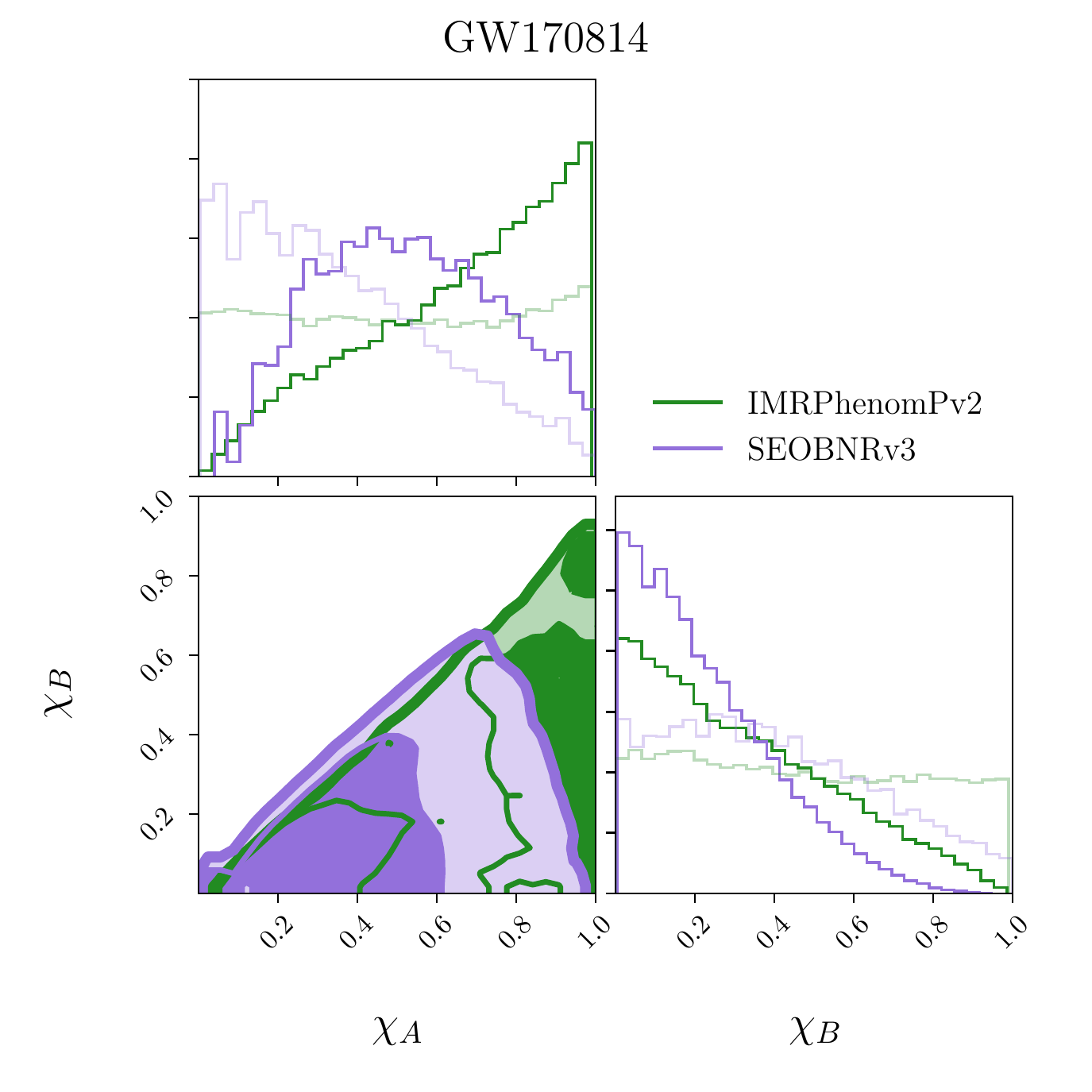}
	\caption{Comparison corner plot for the spin magnitudes for the posteriors obtained using both the IMRPhenomPv2 (green) and SEOBNRv3 (purple) waveforms for GW170814 using the spin sorting. The marginalized posteriors obtained using the mass sorting are shown in lighter green and purple.}
	\label{fig:GW170814}
\end{figure}

For certain sources, we note significant differences in the posteriors obtained with the two different waveform models applied to BBHs in GWTC-1: IMRPhenomPv2~\cite{Husa:2015iqa, Hannam:2013oca, Khan:2015jqa}, which uses an effective precessing spin model, and SEOBNRv3, which uses a fully precessing spin model~\cite{Babak:2016tgq,Taracchini:2013rva, Pan:2013rra}.
For GW150914, the one-dimensional posterior for $\chi_{A}$ is much more tightly constrained for SEOBNRv3, with $\chi_{A} = 0.39^{+0.44}_{-0.24}$ compared to $\chi_{A} = 0.49^{+0.45}_{-0.38}$ for IMRPhenomPv2, a feature which is not as easily recognizable in the $\chi_{1}$ posterior.
A comparison of the spin magnitudes obtained using both waveform models is shown in Fig.~\ref{fig:GW150914}.
Similarly for GW170814, the posterior for $\chi_{A}$ turns over at around $\chi_{A}\sim 0.5$ for SEOBNRv3, but not for IMRPhenomPv2 (Fig.~\ref{fig:GW170814}).
The two-dimensional $\chi_{A}, \chi_{B}$ posterior recovered with SEOBNRv3 is much more tightly clustered around low spins for GW170814 and also for GW170818, although these features are distinguishable in the $\chi_{1}, \chi_{2}$ posteriors as well.

\begin{figure}
	\centering
	\includegraphics[width=0.99\linewidth]{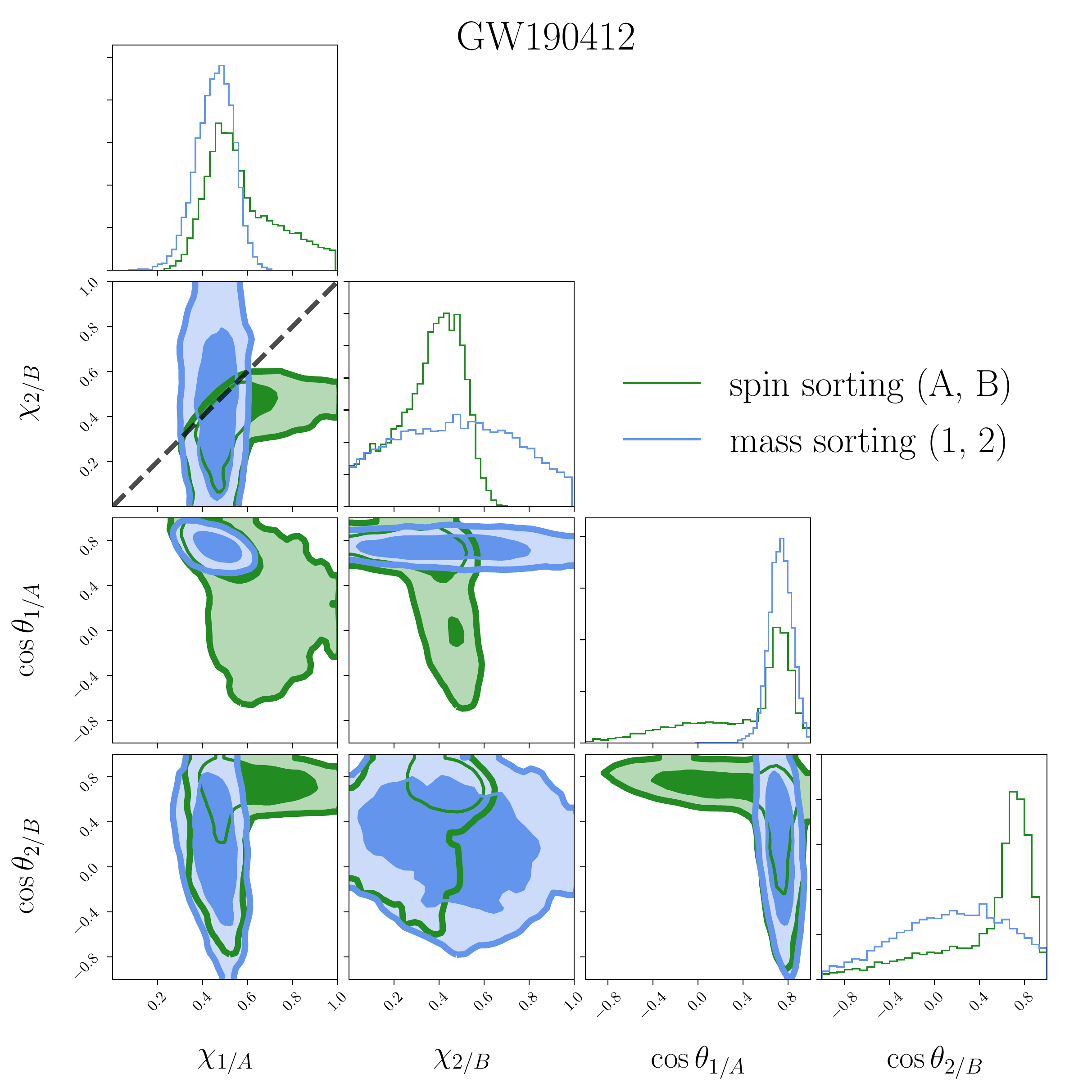}
	\caption{Comparison corner plot for the spin magnitudes and tilts for the posteriors obtained using the SEOBNRv4PHM waveform for GW190412.}
	\label{fig:GW190412}
\end{figure}

\begin{figure}
	\centering
	\includegraphics[width=0.99\linewidth]{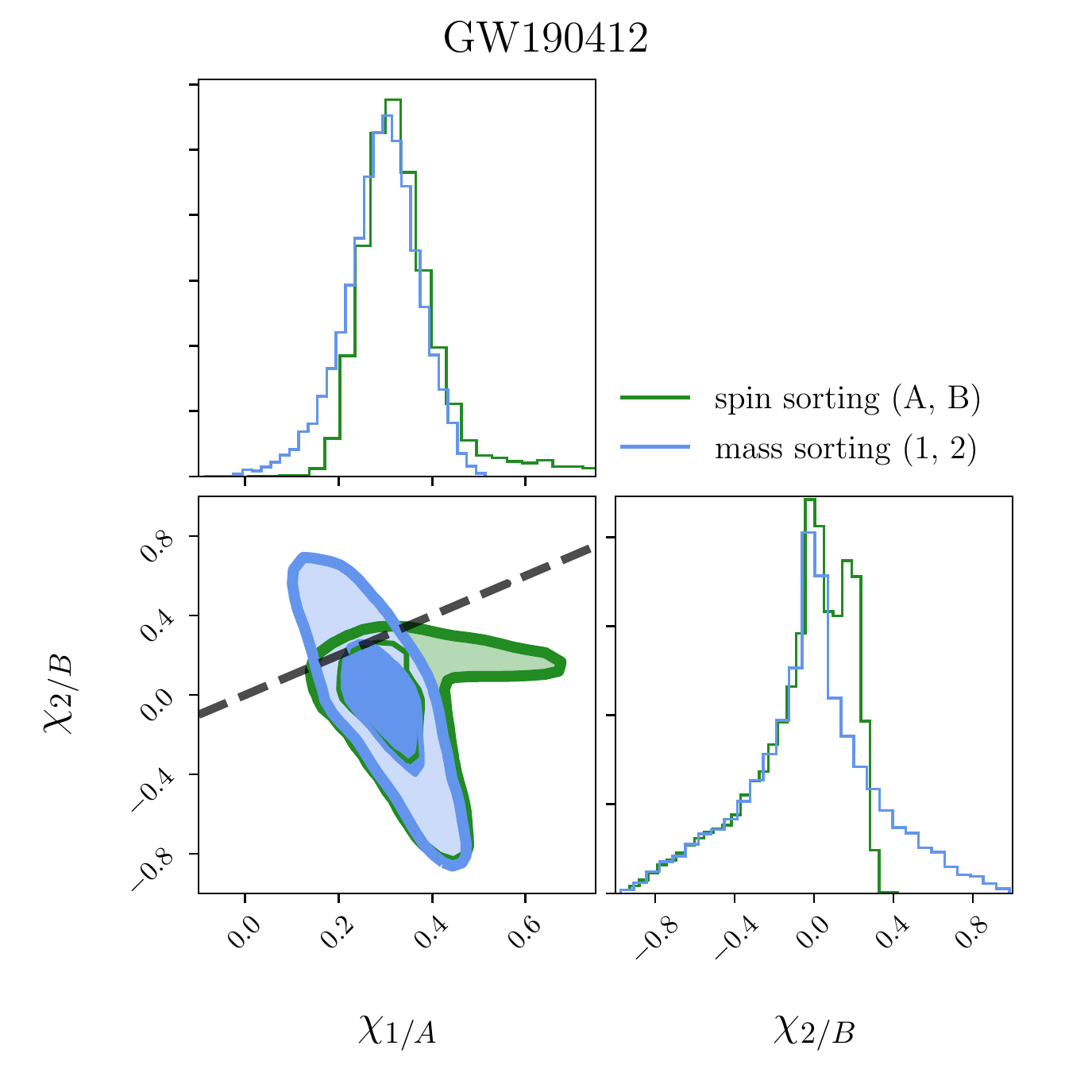}
	\caption{Comparison corner plot for the spin magnitudes for the posteriors obtained using the aligned-spin waveform SEOBNRv4HM\_ROM for GW190412 using both the spin (green) and mass (blue) sorting. Negative values of $\chi$ are included due to anti-alignment.}
	\label{fig:GW190412_aligned}
\end{figure}

For the unequal-mass binary GW190412~\cite{LIGOScientific:2020stg},
the spin sorting introduces degeneracies in the spin parameters that were not present in the mass-based sorting.
The one-dimensional $\chi_{A}$ posterior is much less constrained than $\chi_{1}$, and it features a tail extending to higher spin magnitudes.
Unlike for the other BBH signals, which are consistent with $q=1$, the tilt posteriors for this event change considerably between the mass and spin sortings.
There is a clear degeneracy observed between $\theta_{A}$ and $\theta_{B}$ where either one or the other is constrained to lie in the orbital plane, similar to the pattern observed for $\theta_{1}$ and $\theta_{2}$ in our simulated signal.
The $\chi_{B}$ posterior is more constrained than the $\chi_{2}$ posterior, peaking at $\chi_{B}\sim 0.5$ and ruling out $\chi_{B}\gtrsim 0.7$ with $3\sigma$ credibility across the various waveforms allowing spin precession.
Fig.~\ref{fig:GW190412} compares the posteriors on the spin magnitudes and tilts for both the spin and mass sorting obtained using a waveform model that allows for spin precession.
Based on the application of the spin sorting to this event, we conclude that it is preferable to use the mass sorting when the mass ratio of the binary can be clearly constrained away from equal mass, as expected from our simulations.

When analyzing the posteriors obtained for GW190412 with aligned-spin waveforms, the $\chi_{1}, \chi_{2}$ posterior shown in Fig.~\ref{fig:GW190412_aligned} exhibits a strong correlation depending on orientation with respect to $\vec{L}$: high, aligned primary spins are allowed when the secondary spin is high and anti-aligned; low, aligned primary spins are allowed when the secondary spin is high and aligned.
This correlation, which is due to the strong constraint on the effective aligned spin, $\chi_{\mathrm{eff}}$,
is simply reflected over the $\chi_{A} = \chi_{B}$ boundary when the posteriors are projected into the spin sorting.

For the binary NSs, GW170817~\cite{TheLIGOScientific:2017qsa} and GW190425~\cite{Abbott:2020uma}, two priors were used to include or exclude high spins (maximum $\chi_i$ of 0.99 vs 0.05, respectively), where the high-spin prior allows for the possibility that the binary components are BHs.
For the high-spin prior, the posteriors for both $\chi_{1/A}$ and $\chi_{2/B}$ favor low spin values for both events and for both aligned and precessing-spin waveforms.
The posteriors for the spin magnitudes are less informative for the low-spin prior, since the prior volume is considerably reduced.
For GW190425, we find that the constraints on $\cos{\theta_{A}}$ are tighter than those on $\cos{\theta_{1}}$ for the precessing-spin waveform IMRPhenomPv2\_NRTidal: $\cos{\theta_{A}} = 0.24^{+0.60}_{-0.44}$ compared to $\cos{\theta_{1}} = 0.24^{+0.62}_{-0.65}$ for the high-spin prior, with a similar trend for the low-spin prior. Conversely, the posterior for $\cos{\theta_{B}}$ broadens slightly compared to that of $\cos{\theta_{2}}$, consistent with the behavior observed for the simulated signal.

\begin{figure*}
	\centering
	\includegraphics[width=0.32\linewidth]{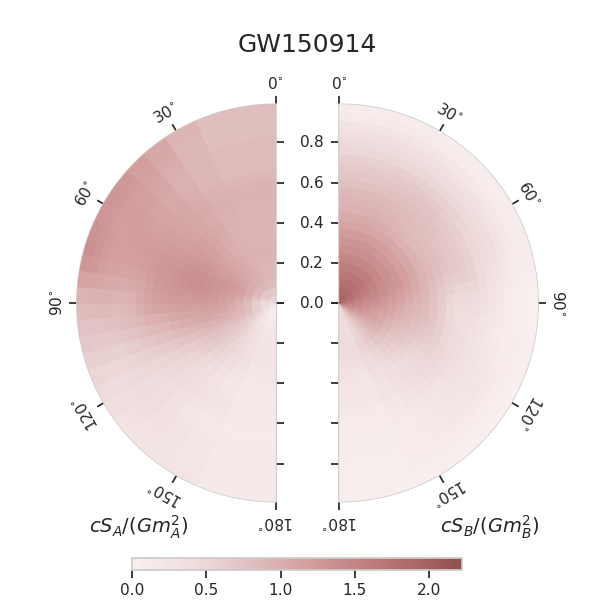}
	\includegraphics[width=0.32\linewidth]{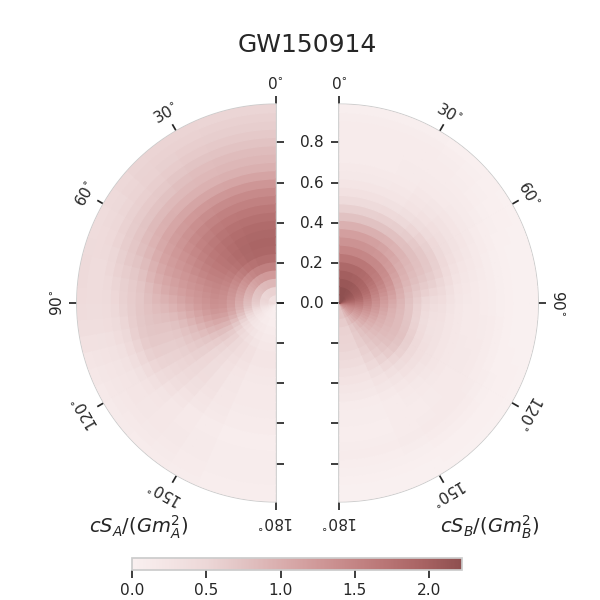}
	\caption{Spin disk plots for the spin-sorted posterior samples for GW150914 using the IMRPhenomPv2 waveform on the left and the SEOBNRv3 waveform on the right.}
	\label{fig:disk_plot}
\end{figure*}

\begin{figure*}
	\centering
	\includegraphics[width=0.32\linewidth]{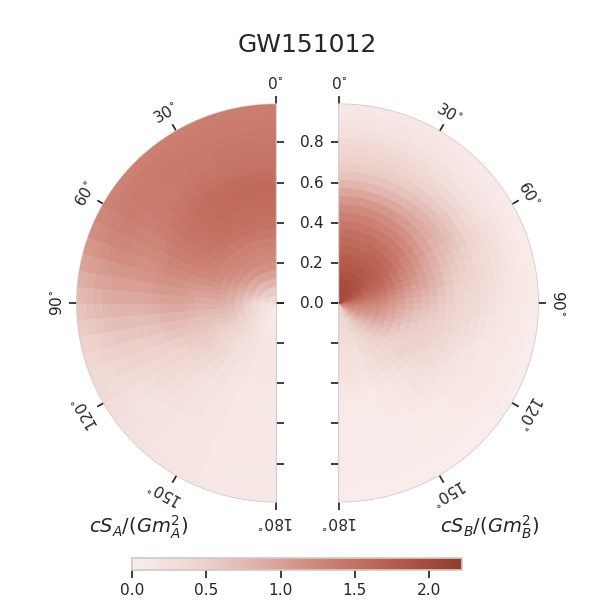}
	\includegraphics[width=0.32\linewidth]{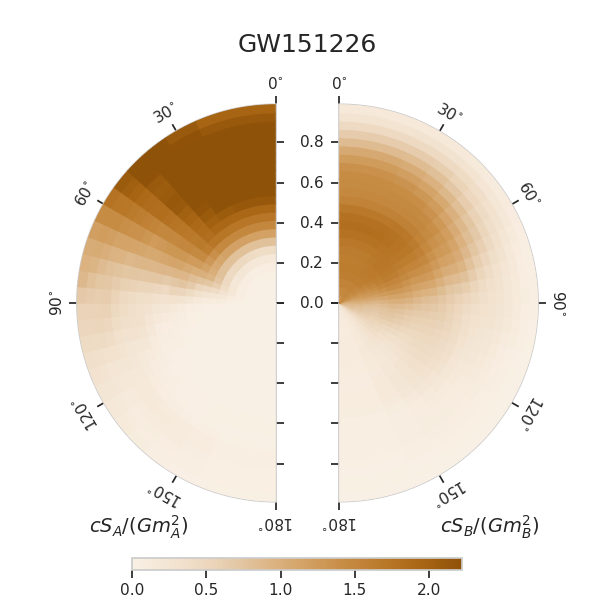}
	\includegraphics[width=0.32\linewidth]{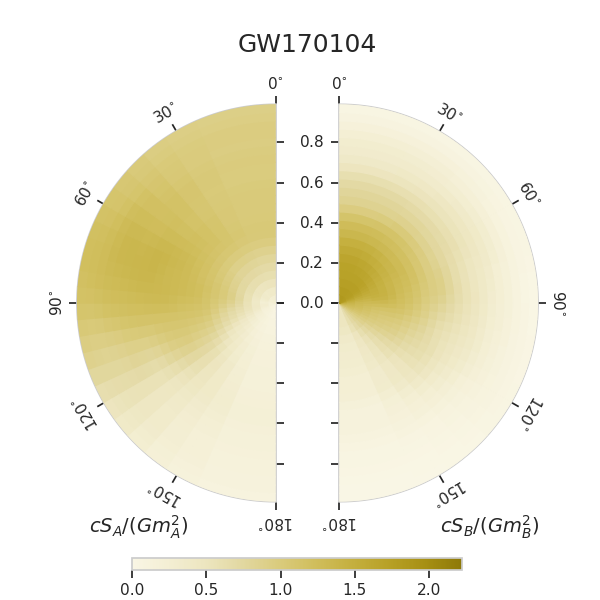}
	\includegraphics[width=0.32\linewidth]{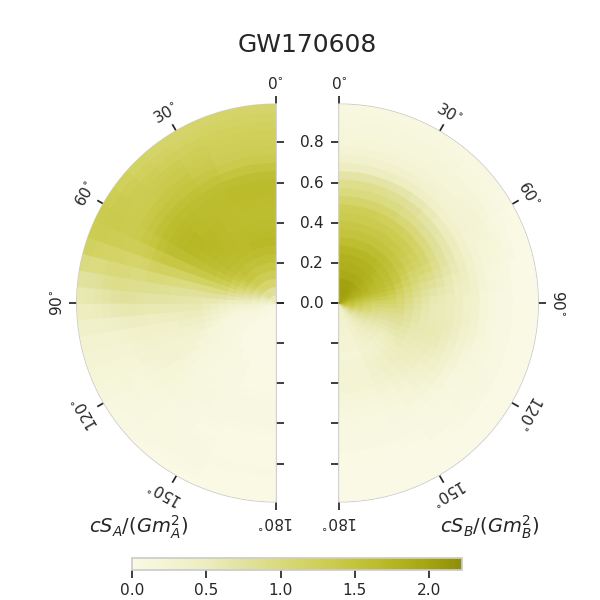}
	\includegraphics[width=0.32\linewidth]{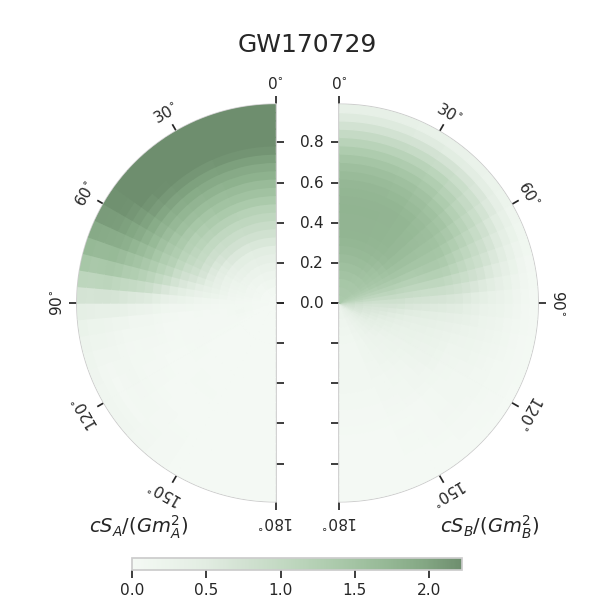}
	\includegraphics[width=0.32\linewidth]{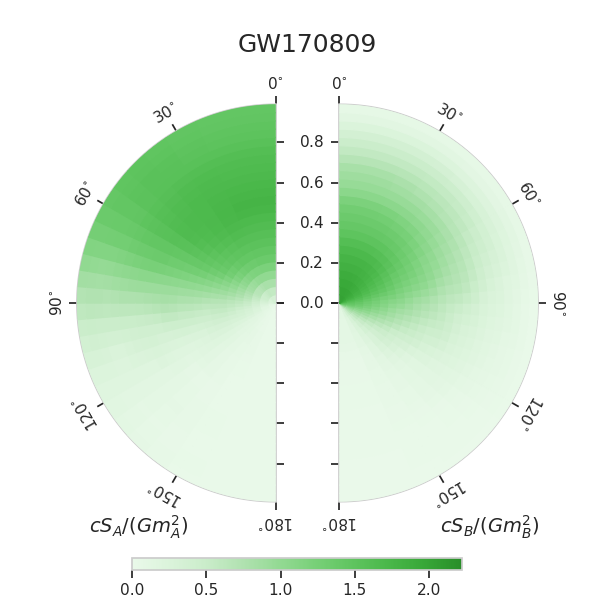}
	\caption{Spin disk plots for the spin-sorted posterior samples for current LVC detections.}
	\label{fig:disk_plot2}
\end{figure*}

\begin{figure*}
	\centering
	\includegraphics[width=0.32\linewidth]{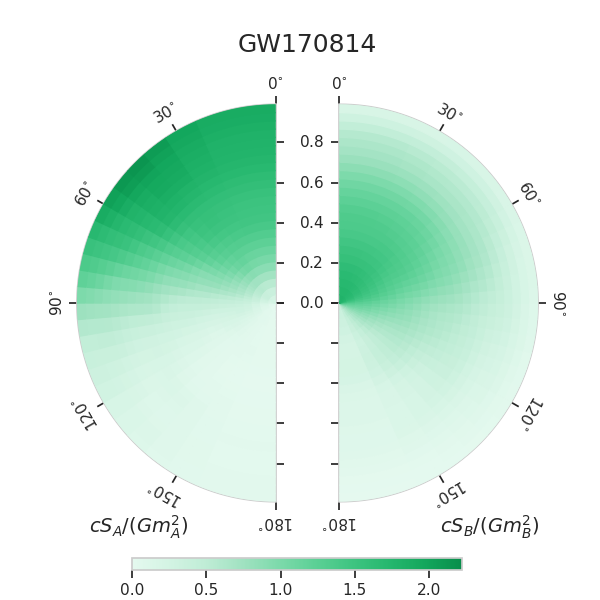}
	\includegraphics[width=0.32\linewidth]{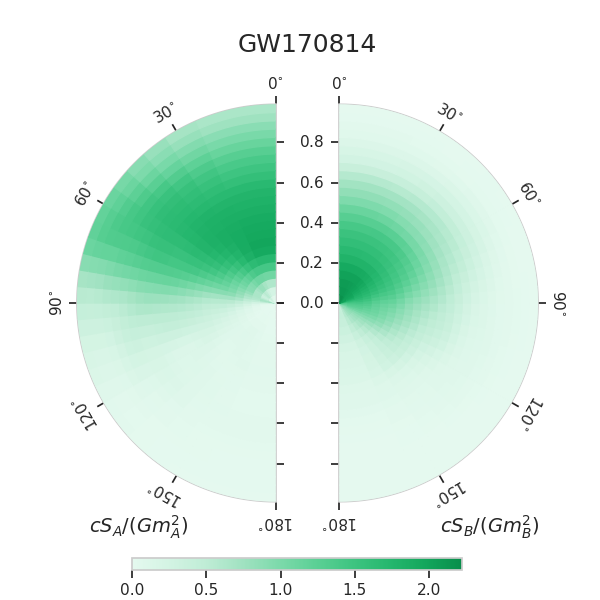}
	\caption{Spin disk plots for the spin-sorted posterior samples for GW170814 using the IMRPhenomPv2 waveform on the left and the SEOBNRv3 waveform on the right.}
	\label{fig:disk_plot3}
\end{figure*}

\begin{figure*}
	\centering
	\includegraphics[width=0.32\linewidth]{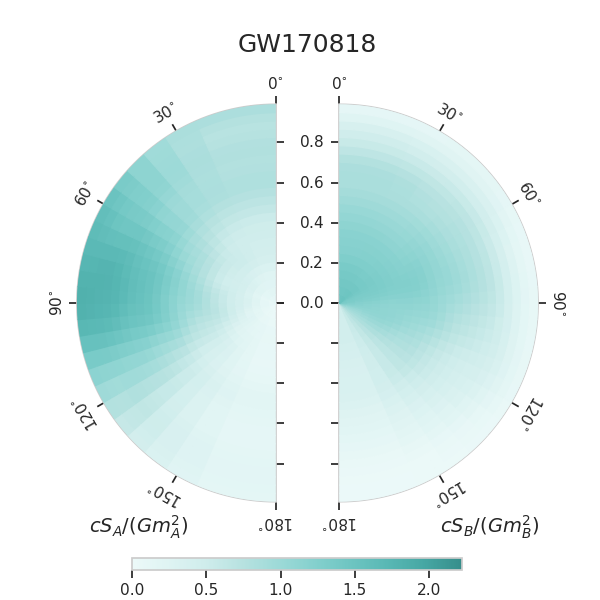}
	\includegraphics[width=0.32\linewidth]{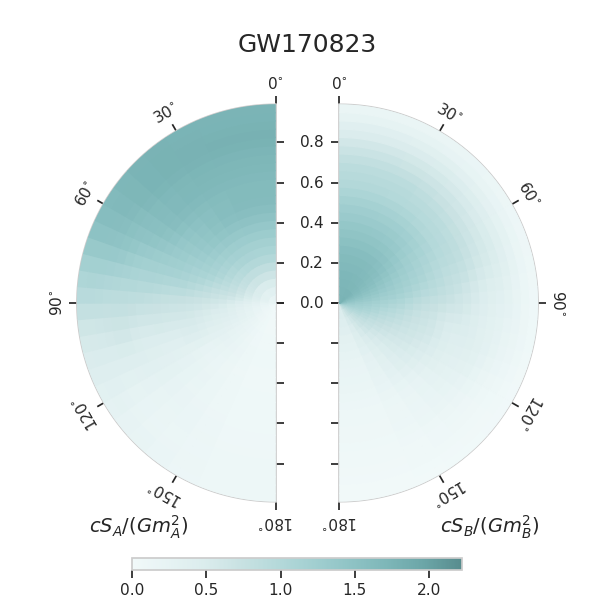}
	\includegraphics[width=0.32\linewidth]{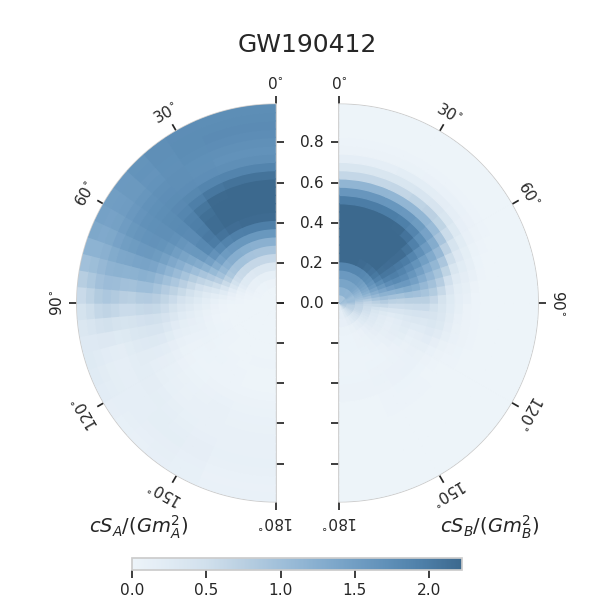}
	\includegraphics[width=0.32\linewidth]{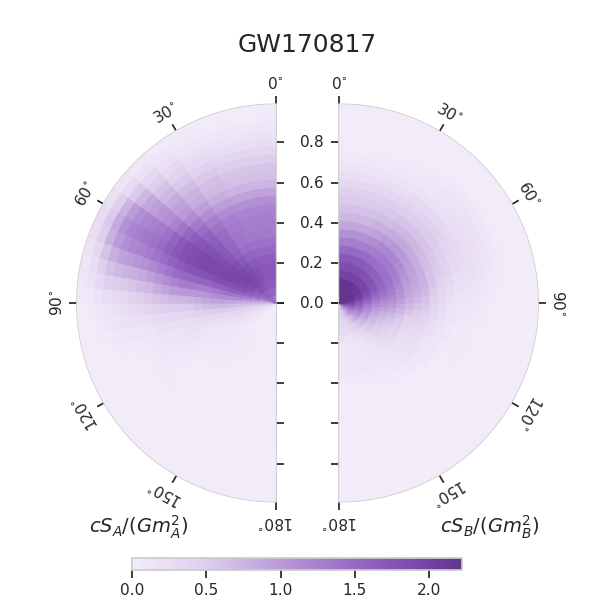}
	\includegraphics[width=0.32\linewidth]{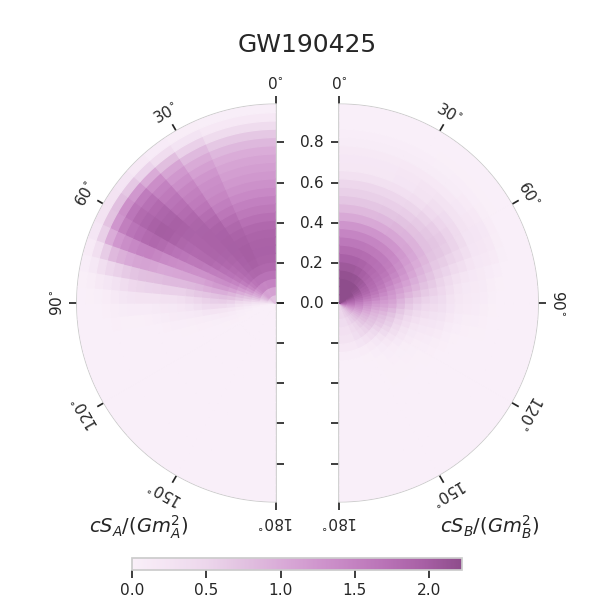}
	\caption{Spin disk plots for the spin-sorted posterior samples for current LVC detections.}
	\label{fig:disk_plot4}
\end{figure*}

\subsection{Spin disk plots}
In Figs.~\ref{fig:disk_plot}--\ref{fig:disk_plot4}, we show the spin disk plots for the spin-sorted posterior samples for the first 13 LVC detections (see e.g. Fig. 5 of \cite{TheLIGOScientific:2016wfe}) calculated using \texttt{PESummary}~\cite{Hoy:2020vys}. The angular direction indicates the misalignment with the orbital angular momentum, while the radial direction shows the spin magnitude for the highest-spinning compact object on the left and the least-spinning object on the right. We show the posterior samples obtained using the IMRPhenomPv2 waveform for BBH events reported in GWTC-1, IMRPhenomPv2\_NRTidal with the high-spin prior for the two BNS detections, and IMRPhenomPv3 for GW190412. We also show the disk plots for the posteriors obtained with the SEOBNRv3 waveform for GW150914 and GW170814 in Figs.~\ref{fig:disk_plot} and \ref{fig:disk_plot3} to supplement the comparison in Figs.~\ref{fig:GW150914} and \ref{fig:GW170814}.

%

\end{document}